\documentclass[paper]{geophysics}
\pdfoutput=1
\usepackage{amsmath}
\usepackage{amssymb}
\usepackage{amsfonts}
\usepackage{color}
\usepackage{array}
\usepackage{seg}

\begin{document}

\renewcommand{\figdir}{.} % figure directory

\title{Improved depth imaging by constrained full-waveform inversion}
\author{Musa Maharramov and Biondo Biondi}
\righthead{CFWI}
\lefthead{Maharramov and Biondi}
\footer{SEP--155}
\maketitle

\begin{abstract}
We propose a formulation of full-wavefield inversion (FWI) as a constrained optimization problem, and describe a computationally efficient technique for solving \emph{constrained full-wavefield inversion} (CFWI). The technique is based on using a total-variation regularization method, with the regularization weighted in favor of constraining deeper subsurface model sections. The method helps to promote ``edge-preserving'' blocky model inversion where fitting the seismic data alone fails to adequately constrain the model. The method is demonstrated on synthetic datasets with added noise, and is shown to enhance the sharpness of the inverted model and correctly reposition mispositioned reflectors by better constraining the velocity model at depth.
\end{abstract}

\section{Introduction}
Full-waveform inversion can achieve high resolution of subsurface velocity reconstruction where the target is shallow and well illuminated by refracted seismic energy from wide-offset surveys \cite[]{SirgueFirstBreak}. However, illumination of deeper targets with refracted energy may require extra wide offset survey acquisitions, or otherwise suffer from poor constraining of deeper model sections. Mathematically, this is a manifestation of the fact that the full-waveform inversion is a mixed-determined problem, with shallow areas of the model overdetermined by the abundance of data, and deeper areas affected by poor resolution and spurious positioning errors. While geological priors such as well tie-ins may provide useful additional constraints in areas of poor illumination, regularization of the nonlinear inversion problems arising in full-waveform inversion is a well-established mathematical technique for dealing with underdetermined problems and noisy data. In particular, it is recognized that the \emph{total variation} (TV) regularization promotes sparsity of model gradients, acting as an ``edge preserving'' constraint complementing or outweighing data fitting in problematic areas \cite[]{anagaw11,guitton12}. However, $\ell_1$ and TV regularized optimization problems are difficult to solve, and the development of efficient numerical solution techniques is a subject of active ongoing research, see e.g. \cite{boyd}.

In this work we proposed a formulation of the full-waveform inversion as a problem of \emph{constrained optimization}, and solve it using the iterative Bregman regularization technique, see e.g. \cite{osher2005}. We demonstrate advantages of the proposed method over unconstrained regularization. The paper concludes with an example of application to the Marmousi synthetic with added noise.  
 
\section{Method}
We begin with the standard formulation of FWI as an unconstrained nonlinear least-squares fitting problem \cite[]{VirieuxOperto}:
\begin{equation}
\begin{aligned}
\|\mathbf{F}(\mathbf{m})\;-\;\mathbf{d}\|_2\;\rightarrow\;\min,
\end{aligned}
\label{eq:fwi}
\end{equation}
where $\mathbf{d}$ is the observed data, $\mathbf{m}$ is the model (i.e., acoustic slowness) and $\mathbf{F}$ is the nonlinear forward modeling operator. Problem (\ref{eq:fwi}) can be solved in either time or frequency domain, with either approach having its advantages \cite[]{VirieuxOperto}. Formulation (\ref{eq:fwi}) equally weights all data, resulting in better illuminated areas being better constrained. Weighted least-squares and priors may help improve recovery of deeper sections, however, the problem still remains underdetermined at greater depths or acquisition blind spots.

In this work we explore an alternative formulation of FWI as a constrained optimization problem,
\begin{equation}
\begin{aligned}
& \|\mathbf{R}\mathbf{m}\|_1\;\rightarrow\; \min,\\
& \mathbf{F}(\mathbf{m})\;=\;\mathbf{d},
\end{aligned}
\label{eq:cfwi1}
\end{equation}
where 
\begin{equation}
\mathbf{R}\mathbf{m}\;=\;|\nabla_{\mathbf{x}} m|,
\label{eq:R}
\end{equation}
is the length of spatial slowness gradient, and $\|\mathbf{R}\mathbf{m}\|_1$ is the \emph{total variation} seminorm of $\mathbf{m}$ \cite[]{Triebel}. Note that in (\ref{eq:cfwi1}) we use an equality constraint for data fitting, which is neither desirable nor realistic in applications to field data. Indeed, formulation (\ref{eq:cfwi1}) is suitable for strictly \emph{underdetermined} problems, and the equality constraint can be enforced only for noise-free data. Since the full-waveform inversion problem is \emph{mixed-determined} and field data are always noisy, we propose to solve the following constrained optimization problem:
\begin{equation}
\begin{aligned}
& \|w(\mathbf{x})\mathbf{R}\mathbf{m}\|_1\;\rightarrow\; \min,\\
& \|\mathbf{F}(\mathbf{m})\;-\;\mathbf{d}\|_2^2\;<\;\sigma^2,
\end{aligned}
\label{eq:cfwi2}
\end{equation}
where $w(\mathbf{x})=w(x^1,x^2,x^3)$ is a weighting function that is above zero only in areas of poor resolution or illumination at depth. In a practical solution algorithm, solving (\ref{eq:cfwi2}) is equivalent to solving (\ref{eq:cfwi1}) with suitable stopping criteria when the desired data misfit $\sigma^2$ is achieved.

We solve (\ref{eq:cfwi2}) using the iterative \emph{Bregman regularization} technique proposed by \cite{osher2005}. Starting from \[\mathbf{m}\;=\;0,\;\;\mathbf{p}_0\;=\;0,\] and given $\mathbf{m}_k$, we iteratively compute slowness $\mathbf{m}_{k+1}$ as the solution of the following unconstrained TV-regularized problem
\begin{equation}
\begin{aligned}
& \lambda \|w(\mathbf{x})\mathbf{R}\mathbf{m}\|_1\;-\;\langle \mathbf{p}_k,\mathbf{m}-\mathbf{m}_k \rangle + \|\mathbf{F}(\mathbf{m})-\mathbf{d}\|_2^2\;\rightarrow\; \min,
\label{eq:utvfwi}
\end{aligned}
\end{equation}
where an element $\mathbf{p}_{k+1}$ of the \emph{subgradient} of $\lambda \|w(\mathbf{x})\mathbf{R}\mathbf{m}\|_1$ is computed as
\begin{equation}
\begin{aligned}
\mathbf{p}_{k+1}\;=\;\mathbf{p}_k\;-\left.\;\nabla_{\mathbf{m}} \|\mathbf{F}(\mathbf{m})-\mathbf{d}\|_2^2 \right\vert_{\mathbf{m}=\mathbf{m}_{k+1}}.
\end{aligned}
\label{eq:sub}
\end{equation}
Note that (\ref{eq:utvfwi}) describes a TV-regularized inversion \cite[]{ROF} that may yield an edge-preserving or ``blocky'' approximation to the solution of the full-waveform inversion problem (\ref{eq:fwi}). The first two terms in (\ref{eq:utvfwi}) are known as the \emph{Bregman distance} \cite[]{Bregman1967}, that is equivalent to extracting an approximately quadratic function from the regularization term. We use the regularization approach based on solving the single unconstrained problem similar to (\ref{eq:utvfwi}) with $\mathbf{p}_k\equiv 0$ in our work on TV-regularized time-lapse FWI \cite[]{musasep1551}. It is important to note, however, that in this work, solution of an unconstrained TV-regularized problem (\ref{eq:utvfwi}) is just a single iteration of an algorithm for solving the \emph{constrained} problem (\ref{eq:cfwi2}). We solve problem (\ref{eq:utvfwi}) using nonlinear conjugate gradients (NCG) algorithm \cite[]{Nocedal}, with the smoothed TV regularization term
\begin{equation}
\| \sqrt{|\nabla_{\mathbf{x}} m|^2+\epsilon} \|_1,
\label{eq:regtv}
\end{equation}
where $\epsilon\approx 10^{-5}$ is chosen as a threshold for realistic values of the slowness.

Choice of the regularization parameter $\lambda$ in (\ref{eq:utvfwi}) can be based on achieving better conditioning of problem (\ref{eq:utvfwi}), and unlike the traditional penalty function/continuation methods for solving (\ref{eq:cfwi1}), $\lambda$ does not increase between iterations \cite[]{goldstein,cai}. Furthermore, iterations (\ref{eq:utvfwi},\ref{eq:sub}) can be shown to converge to a solution  of (\ref{eq:cfwi1}) (or a solution of (\ref{eq:cfwi2}) for some $\sigma>0$ for noisy data) regardless of the value of $\lambda>0$, as can be demonstrated by a trivial extension of the technique of \cite{cai}. However, the rate of convergence does depend on the value of the regularization parameter $\lambda$, making the application of Bregman regularization to some nonlinear operators $\mathbf{F}$ problematic. However, our experiments indicate that the value of the regularization parameter chosen to improve the convergence of NCG for (\ref{eq:utvfwi}) results in good overall convergence of Bregman iterations.

Iterations are stopped when the data misfit reaches a desired value of $\sigma>0$ \cite[]{osher2005,cai}. In practical applications where $\sigma$ may not be known a-priori, iterations may continue until the effects of overfitting start exceeding the edge-preserving effects of the regularization term (\ref{eq:regtv}). Note that instead of using the NCG to solve (\ref{eq:utvfwi}) with the smoothed regularization term (\ref{eq:regtv}), problem  (\ref{eq:utvfwi})  can be solved using \emph{split Bregman} method that only requires iterative solution of nonlinear least squares problems and soft thresholding \cite[]{goldstein}. However, our numerical experiments indicate that the NCG applied to the smoothed (\ref{eq:regtv}) has equivalent performance and accuracy. 

\section{Numerical examples}
We apply the method to the synthetic dataset used in \cite[]{musasep1551}, generated for the Marmousi velocity model over a 384$\times$122 grid with a 24 m grid spacing. The inversion is carried out in the frequency domain for 3.0, 3.6, 4.3, 5.1, 6.2, 7.5, 9.0, 10.8, 12.8, and 15.5 Hz with time-domain forward modeling \cite[]{SirgueFirstBreak}. The frequencies are chosen based on the estimated offset to depth range of the data \cite[]{SirguePratt}. The acquisition has 192 shots at a depth of 16 m with a 48 m spacing, and 381 receivers at a depth of 15 m with a 24 m spacing. The minimum offset is 48 m. The source function is a Ricker wavelet centered at 10.1 Hz. Absorbing boundary conditions are applied along the entire model boundary, including the surface (thus suppressing multiples). A smoothed true model shown in \cite[]{musasep1551} is used as a starting model for the inversion. The smoothing is performed using a triangular filter with a 20-sample half-window in both vertical and horizontal directions. Random Gaussian noise is added to the noise-free synthetic data to produce a noisy dataset with 7 dB signal-to-noise ratio. The result of model inversion from the 7 dB SNR synthetic data is shown in Figure~\ref{fig:base}.  Up to 10 iterations of the nonlinear conjugate gradients algorithm \cite[]{Nocedal} are performed for each frequency. Neither regularization nor model priors are used.
\plot{base}{width=\columnwidth}
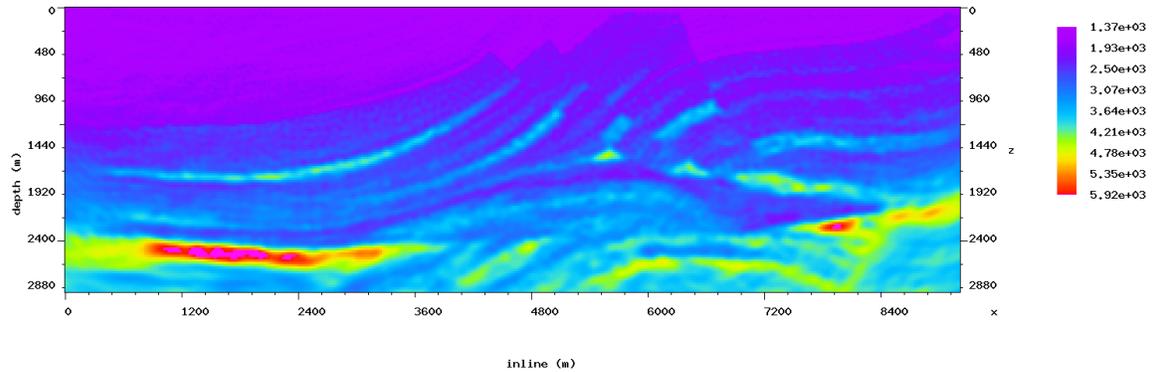
{Inversion of 7dB SnR synthetic using the unregularized FWI, 10 iterations per frequency.}
Figure~\ref{fig:cfwi} shows the results of solving the proposed constrained FWI (\ref{eq:cfwi2}). Only 5 NCG iterations were used for solving each problem (\ref{eq:utvfwi}), with only two outer (Bregman) iterations, resulting in roughly the same compute time as in our standard FWI experiment shown in Figure~\ref{fig:base} (10 gradient evaluations using the adjoint state method). The weighting function $w(\mathbf{x})$ was set to 1 below 2100 m and zero above 2000 m, thus the regularization is applied to less constrained areas.
\plot{cfwi}{width=\columnwidth}
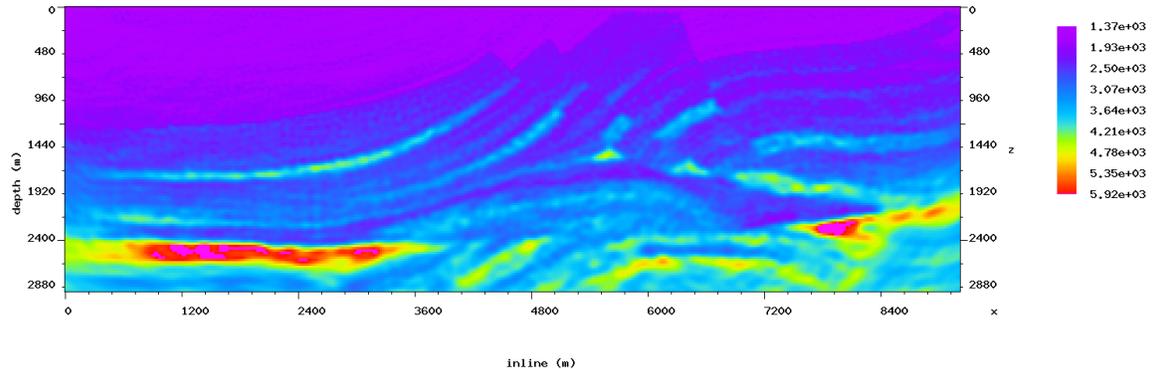
{Inversion of 7dB SnR synthetic by solving constrained TV-regularized FWI (\ref{eq:cfwi2}) using Bregman iterative procedure, 10 iterations per frequency. Note that while the shallow parts are similar to Figure~\ref{fig:base}, deeper sections below 2km are more focused, and the poorly illuminated and mispositioned intervals in the left part of the model have been improved. }
Our results in Figure~\ref{fig:cfwi} indicate that CFWI has improved the deepest section of the model while matching the standard FWI in more shallow well-constrained areas. The result of Figure~\ref{fig:cfwi} is closer to the clean synthetic inversion shown in \cite[]{musasep1551}, and has better delineated interfaces.  

\section{Conclusions}

We have proposed a new formulation of FWI as a constrained optimization problem (CFWI), and demonstrated the CFWI to be a viable technique for improving depth resolution and accuracy of FWI. Application of Bregman iterative regularization provides a computationally efficient solution method for CFWI that can be easily built on top of the existing solvers. Application of CFWI to field data will be the subject of future work. 

\section{Acknowledgments}

The authors would like to thank Jon Claerbout and Stewart Levin for a number of useful discussions, and the Stanford Center for Computational Earth and Environmental Sciences for providing computing resources.

\bibliographystyle{seg}  
\bibliography{mmbbcfwi}

\end{document}